\documentclass[number,citesort,MSNbibl,dvips]{arxbj}
\usepackage{mathrsfs,algorithm}
\usepackage{graphicx}

\aid{0}
\volume{18}
\issue{2}
\pubyear{2012}
\firstpage{496}
\lastpage{519}
\doi{10.3150/10-BEJ345}

\makeatletter

\newcommand{\eqref}[1]{(\ref{#1})}

\renewcommand{\mathcal}{\mathscr}

\renewcommand{\epsilon}{\varepsilon}
\newcommand{\Ran}{\operatorname{Ran}}
\newcommand{\expec}{\mathrm{E}}
\newcommand{\var}{\operatorname{Var}}
\newtheorem{theorem}{Theorem}

\newtheorem{lemma}{Lemma}
\makeatother

\begin{document}
\begin{frontmatter}

\title{Bayesian estimation of a bivariate copula using the Jeffreys prior}
\runtitle{Bayesian estimation of a bivariate copula}

\begin{aug}
\author[1]{\fnms{Simon} \snm{Guillotte}\thanksref{1}\ead[label=e1]{simon.guillotte@gmail.com}}
\and
\author[2]{\fnms{Fran\c{c}ois} \snm{Perron}\corref{}\thanksref{2}\ead[label=e2]{perronf@dms.umontreal.ca}}
\runauthor{S. Guillotte and F. Perron}
\address[1]{Department of Mathematics and
Statistics, University of Prince Edward Island, 550 University Ave.,
Charlottetown,
Prince Edward Island, Canada C1A 4P3.\\ \printead{e1}}
\address[2]{D\'{e}partement de Math\'{e}matiques et
Statistique de l'Universit\'{e} de Montr\'{e}al, Pavillon
Andr\'{e}-Aisenstadt, 2920 Chemin de la Tour, Montr\'{e}al, Qu\'{e}bec,
Canada H3C 3J7.\\ \printead{e2}}
\end{aug}

\received{\smonth{1} \syear{2010}}
\revised{\smonth{8} \syear{2010}}

%
\begin{abstract}
A bivariate distribution with continuous margins can
be uniquely decomposed via a copula and its marginal distributions.
We consider the problem of estimating the copula function and adopt a Bayesian
approach. On the space of copula functions, we construct a finite-dimensional
approximation subspace that is parametrized by a doubly stochastic
matrix. A
major problem here is the selection of a prior distribution on the
space of
doubly stochastic matrices also known as the Birkhoff polytope. The main
contributions of this paper are the derivation of a
simple formula for the Jeffreys prior and showing that it is
proper. It is known in the literature that for a complex problem
like the one treated here, the above results are difficult to obtain.
The Bayes
estimator resulting from the Jeffreys prior is then evaluated
numerically via
Markov chain Monte Carlo methodology. A rather extensive simulation experiment
is carried out. In many cases, the results favour the Bayes estimator
over frequentist estimators such as the standard kernel estimator and Deheuvels'
estimator in terms of mean integrated squared error.
\end{abstract}

%

\end{frontmatter}
%

\section{Introduction}\label{intro}

Copulas have received considerable attention recently because of their
increasing use in multiple fields such as environmental
studies, genetics and data networks. They are also currently very
popular in quantitative finance and insurance; see
Genest \textit{et al.} \cite{Genest3}. Since it is precisely the
copula that describes the dependence
structure among various random quantities, estimating a~copula is part
of many
techniques employed in these fields. For instance, in risk measurement, the
\textit{value at risk (VaR)} is computed by simulating asset log
returns from
a fitted joint distribution, for which the dependence structure between
the assets is modelled by a copula. Further financial examples in which
copulas are estimated
are provided in Embrechts \textit{et al.} \cite{Embrechts03} and
the books written by Cherubini~\textit{et~al.}~\cite{Cherubini1},
McNeil \textit{et al.} \cite{McNeil05} and Trivedi and Zimmer
\cite{Trivedi1}.
In this paper, we provide new generic methodology for estimating
copulas within a Bayesian framework.

Let us first recall that a bivariate copula $C$ is a cumulative
distribution function on
$S=[0,1]\times[0,1]$ with uniform margins. In this paper, we denote
the space of all copulas by $\mathcal C$. Every $C\in\mathcal C$ is Lipschitz
continuous, with a common Lipschitz constant equal to one:
\begin{equation}\label{Lip}
|C(u_1,v_1)-C(u_2,v_2)| \leq|u_1-u_2|+|v_1-v_2|\qquad \mbox{for all } (u_i,v_i) \in S, i\in\{1,2\}.
\end{equation}

The space $\mathcal C$ is bounded above and below by the so-called Fr\'{e}chet--Hoeffding
copulas, that is, for every $C \in\mathcal C$,
\[
\max(0,u+v-1)\leq C(u,v) \leq\min(u,v)\qquad \mbox{for all } (u,v)
\in S.
\]
Sklar's theorem states that a bivariate cumulative distribution
function $F$ is completely
characterized by its marginal cumulative distribution functions $F_X,
F_Y$ and its copula $C$. More
precisely, we have the representation
\begin{equation}
F(x,y) = C(F_X(x),F_Y(y))\qquad \mbox{for all } (x,y) \in\mathbb{R}^2,
\label{Sklar}
\end{equation}
where $C$ is well defined on $\Ran(F_X)\times\Ran(F_Y)$; see
Nelsen \cite{Nelsen1}. In particular, the copula is
unique if $F_X$ and $F_Y$ are
continuous and, in this case, we have the following expression for the
copula:
\begin{equation}
C(u,v)=F(F_X^{-1}(u),F_Y^{-1}(v))\qquad \mbox{for all } (u,v) \in S.
\label{Copula}
\end{equation}

Let $\{(x_i,y_i), i=1,\ldots,n\}$ be a sample where every $(x_i,y_i)$
is a
realization of the random couple $(X_i,Y_i)$, $i=1,\ldots,n$, with joint
cumulative distribution function $F$, and continuous marginal
cumulative distribution functions
$F_X$ and $F_Y$. We consider the problem of estimating the unknown
copula $C$ of $F$ by a
copula $\hat C$, where $\hat C$ depends on the sample. In this problem, the
individual marginal distributions are treated as nuisance parameters. The
literature presents three generic approaches for estimating $C$, namely the
fully parametric, the semi-parametric and the nonparametric approaches.
Below, we briefly describe each approach and focus on two
nonparametric estimators, since we will subsequently compare our
estimator with
these.

\textit{The fully parametric approach.} In this framework, parametric
models are assumed
for both the marginal distribution functions $F_X$ and $F_Y$ and for
the copula
$C$. See Joe \cite{Joe1}; Cherubini~\textit{et al.}
\cite{Cherubini1}, Joe \cite{Joe2}, and, in a
Bayesian setup, Silva and Lopes \cite{Silva1}.

\textit{The semi-parametric approach.} Here, a parametric model is
assumed only for the copula function $C$, not for the margins. In this
setup, Genest \textit{et al.} \cite{Genest2} have proposed the use
of rescaled empirical
distribution functions, such as the estimates $\hat F_X$ and $\hat F_Y$
and a pseudo-likelihood estimator for $C$. The authors show that the
resulting estimator is consistent
and asymptotically normal. In Kim \textit{et al.} \cite{Kim1},
comparisons are made
between the
fully parametric approach and the semi-parametric approach proposed by
Genest \textit{et al.} \cite{Genest2}. More recently, in a Bayesian
setup, Hoff \cite{Hoff1}
proposes a
general estimation procedure, via a likelihood based on ranks, that
does not
depend on any parameters describing the marginal distributions. The latter
methodology can accommodate both continuous and discrete data.

\textit{The nonparametric approach.} This approach exploits
equation (\ref{Copula}). Here, we describe Deheuvels' estimator and
the kernel
estimator. Let $\hat F$ be the empirical cumulative distribution
function, and
let $\hat F_X^{-1}$ and $\hat F_Y^{-1}$ be the generalized inverses of its
marginal cumulative distribution functions. Any copula
$\hat C \in\mathcal{C}$ is said to satisfy the Deheuvels constraint
associated with $\hat F$ provided that for all
$i,j=1,\ldots,n$,
\begin{eqnarray*}
\hat C({i/n},{j/n})&=&\hat F\bigl(\hat
F_X^{-1}({i/n}),\hat F_Y^{-1}({j/n})\bigr) \\
&=& (1/n) \sum_{k=1}^n \mathbf{1}\bigl(\operatorname{rank}(x_k)\leq i, \operatorname{rank}(y_k)\leq
j\bigr)\qquad\mbox{(Deheuvels' constraint). }
\end{eqnarray*}
In Deheuvels \cite{Deheuvels1}, the asymptotic
behaviour of the class of copulas
$\mathcal{C}_{\hat F}\subset\mathcal{C}$
satisfying the Deheuvels constraint associated with $\hat F$ is
described. Note that, in the literature, the so-called \textit
{empirical copula}
$\hat{C}_{\mathrm{emp}}(u,v)= \hat F(\hat F_X^{-1}(u),\hat F_Y^{-1}(v))$, for
all $(u,v)\in S,$ is a~function that satisfies the Deheuvels constraint
and is
often used as an estimator for $C$ even though it is not a genuine
copula. In
Lemma \ref{Deheuvelsestimator}, we propose an estimator $\hat{C}_{\mathrm{DEH}}$
that satisfies Deheuvels' constraint and which, unlike $\hat
{C}_{\mathrm{emp}}$, is itself a
copula so that~\mbox{$\hat{C}_{\mathrm{DEH}}\in\mathcal{C}_{\hat F}$}. This
estimator, which we
call Deheuvels' estimator henceforth, is based on
ranks. One nice property of rank-based estimators is their invariance under
strictly increasing transformations of the margins. Therefore, if
$\varphi$ and
$\psi$ are two strictly increasing functions, then the Deheuvels
estimator based on
the original sample and the one based on the sample
$\{(\varphi(x_i),\psi(y_i))\dvt i=1,\ldots,n\}$ are identical.
This is a
desirable property for a copula estimator since it is inherent to copulas
themselves.

Moreover, if $\hat F$ is a smooth kernel estimator of $F$ ($\hat F_X$
and $\hat F_Y$ are continuous say), then
\begin{equation}\label{Gaussker}
\hat C(u,v)= \hat F(\hat F_X^{-1}(u),\hat F_Y^{-1}(v))\qquad\mbox{for all } (u,v)\in S,
\end{equation}
is called a kernel estimator for $C$, and we have $\hat{C} \in
\mathcal{C}$. Asymptotic properties of such
estimators are discussed in Fermanian and Scaillet \cite{Fermanian1}, and the reader is
referred to
Charpentier~\textit{et al.} \cite{Charpentier1} for a recent
review. In particular, the so-called
Gaussian kernel estimator is given by~\eqref{Gaussker} using
\[
\hat{F}(x,y)=(1/n)\sum_{i=1}^n\Phi\biggl(\frac{x-x_i}{h}\biggr)\Phi
\biggl(\frac{y-y_i}{h}\biggr)\qquad\mbox{for all } x,y,
\]
where $\Phi$ denotes the standard univariate Gaussian cumulative
distribution function and $h>0$ is the value of the bandwidth.

Both of the nonparametric estimators discussed above have good asymptotic
properties. On the other hand, they may not be optimal for small sample sizes.
This could be an inconvenience when working with small samples, and we think
practitioners should be aware of this. We illustrate some of these
situations by
a simulation study in Section~\ref{sec5}.\looseness=1

Our aim is to develop a Bayesian alternative for the estimation of
$C$ that circumvents this problem. Following Genest \textit{et al.}
\cite{Genest2}, when the marginal
distributions are unknown, we use rescaled empirical distribution
functions as their estimates. In view of this, our methodology can be
called empirical Bayes. When the marginal distributions are known, the
sample $\{(x_i,y_i),i=1,\ldots,n\}$ is replaced by $\{
(F_X(x_i),F_Y(y_i)),i=1,\ldots,n
\}$, which is a sample from the uniform distribution $\mathcal{U}(0,1)$. In this case, our
procedure is purely Bayesian. In both cases, our estimator has the
property of
being invariant under monotone transformations of the margins, just like
Deheuvels' estimator.

Our model is obtained as follows. First, in Section \ref{sec2} we construct
an approximation subspace $\mathcal{C}^*\subset\mathcal{C}$. This is
achieved by considering the sup-norm $\|\cdot\|_{\infty}$
and setting a~precision $\epsilon> 0$ so that for every copula $C \in
\mathcal{C}$ there exists a copula $C^* \in\mathcal{C}^*$ such that
$\|C^*-C\|_{\infty} \leq
\epsilon$. Moreover, $\mathcal{C}^*$ is finite dimensional; it is
parametrized by a doubly stochastic matrix $P$. Then, our estimator
$\hat C$ is
obtained by concentrating a prior on $\mathcal{C}^*$ and computing the
posterior mean, that is, the Bayes estimator under squared error loss.
Now two problems arise, the first one is the prior selection on
$\mathcal{C}^*$
and the second one concerns the numerical evaluation of the Bayes estimator.
These are the topics of Sections \ref{sec3} and \ref{sec4}, respectively. While the
problem of
evaluating the Bayes estimator is solved using a Metropolis-within-Gibbs
algorithm, the choice of the prior distribution is a~much more delicate problem.
A copula from our model can be written as a~finite
mixture of cumulative distribution functions. The mixing weights form a
matrix $W$
that is proportional to a doubly stochastic matrix. Therefore,
specifying a prior on $\mathcal{C}^*$ boils down to specifying a prior
for the mixing
weights. We assume that we do not have any information that we could
use for
the construction of a subjective prior. It is not our intention to obtain
a Bayes estimator better than some other given estimator. For these
reasons we
shall rely on an objective prior, and a natural candidate is the
Jeffreys prior.
The main contributions of our paper are the derivation of a simple expression
for the Jeffreys prior and showing that it is proper. The fact that
these results are generally difficult to come up with, for finite mixture
problems, has been raised before in the literature; see Titterington \textit{et al.} \cite{Titterington85}
and Bernardo and Gir\'{o}n \cite{Bernardo88}. Moreover, here we face the
additional difficulty that the mixing weights are further constrained, since
their sum is fixed along the rows and the columns of $W$. To the best
of our
knowledge, nothing has yet been published on this problem. In Section
\ref{sec5}, we report the results of an extensive simulation study in which we
compare our
estimator with Deheuvels' estimator and the Gaussian kernel estimator.
Finally, a~discussion
is provided in Section \ref{sec6} to conclude the paper.

\section{The model for the copula function}\label{sec2}

For every $m > 1$, we construct a finite-dimensional
approximation subspace $\mathcal{C}_m \subset\mathcal{C}$. The
construction of
$\mathcal{C}_m$ uses a basis that forms a partition of unity.
A partition of unity is a~set of nonnegative functions
$g=\{g_i\}_{i=1}^m,$ such
that $mg_i$ is a probability density function on $[0,1]$ for all $
i=1,\ldots,m,$ and
\[
\sum_{i=1}^mg_i(u)=1\qquad \mbox{for all } u\in[0,1].
\]
Particular examples are given by indicator functions
\begin{equation}\label{c}
\cases{
g_1=\mathbf{1}_{[0,1/m]}, \cr
g_i=\mathbf{1}_{((i-1)/m,i/m]}, &\quad  $i=2,\ldots,m,$
}
\end{equation}
and Bernstein polynomials
\begin{equation}\label{b}
g_i=B_{i-1}^{m-1},\qquad  i=1,\ldots,m,
\end{equation}
where
\[
B_i^m(u)={m\choose i}
u^i(1-u)^{m-i}\qquad \mbox{for all } u\in[0,1].
\]
See Li \textit{et al.} \cite{Li1} for more examples of partitions
of unity. In the
following, let
$G=(G_1,\ldots,G_m)^{\top}$, where $G_i(u)=\int_0^ug_i(t)\,\mathrm{d}t$, for all $u\in[0,1]$, $i=1,\ldots,m$, and let
\begin{equation}
C_P^*(u,v)=mG(u)^{\top}PG(v)\qquad \mbox{for all } (u,v) \in
S,\label{CP*}
\end{equation}
where $P$ is an ${m\times m}$ doubly stochastic matrix. The following
lemma is
straightforward to prove.
\begin{lemma}
For every doubly stochastic matrix $P$, $C_P^*$ is an absolutely continuous
copula.
\end{lemma}
%

For a fixed partition of unity, we now define the approximation space as
\[
\mathcal{C}_m=\{C_P^*\dvtx P \mbox{ is an $m\times m$ doubly stochastic
matrix}\}.
\]
The
approximation order of $\mathcal{C}_m$ is now discussed. It depends on the
choice of the basis~$G$. Let $\mathcal{G}_m=
\{(i/m,j/m)\dvt i,j=1,\ldots,m\}$ be a uniformly spaced grid on the unit square~$S$.
For a given copula~$C$, let $R_C=(C(i/m,j/m))_{i,j=1}^m$ be the
restriction of $C$ on~$\mathcal{G}_m$. Let
\[
D=\pmatrix{
1 & 0 & 0 & 0 & \cdots& 0 \cr
-1 & 1 & 0 & 0 & \cdots& 0 \cr
0 & -1 & 1 & 0 & \cdots& 0 \cr
0 & 0 & -1 & 1 & \cdots& \vdots\cr
\vdots& \vdots& \vdots& \vdots& \ddots& 0 \cr
0 & 0 & 0 & 0 & -1 & 1
}.
\]
Then $P_C=mDR_CD^{\top}$ is a doubly stochastic matrix. Upper
bounds for $\|C_{P_C}^*-C\|_{\infty}$ are given in the following lemma.

\begin{lemma}\label{copulaapprox}Let $C$ be a copula and let
$C^*=C_{P_C}^*\in\mathcal{C}_m$, where $C_{P_C}^*$ is obtained
by~\eqref{CP*}.
\begin{longlist}
\item[(a)] For a model using indicator functions basis~\eqref{c}, we have
$R_{C^*}=R_C$
and
$\|C^*-C\|_{\infty}\leq2/m$.
\item[(b)] For a model using the Bernstein basis~\eqref{b}, we have
$\|C^*-C\|_{\infty}\leq1/\sqrt{m}$.
\end{longlist}
\end{lemma}

\begin{pf}
(a) A direct evaluation shows that
$R_{C^*}=R_C$. From the Lipschitz condition~\eqref{Lip}, if two
copulas $C_1$
and $C_2$ satisfy
the constraint $R_{C_1}=R_{C_2}$, then
$\|C_1-C_2\|_{\infty}\leq2/m$.

(b) First, it is well known that
$mG^{\top}D=(B_1^m,\ldots,B_m^m)$. For any $(u,v)\in S$, consider two
independent random variables, $X$ and $Y$, where
$X$ follows a $\operatorname{Binomial}(m,u)$ distribution and $Y$ follows a
$\operatorname{Binomial}(m,v)$ distribution. We have
\[
C^*(u,v)=\expec_{u,v}[C(X/m,Y/m)].
\]
Therefore,
\begin{eqnarray*}
\sup_{(u,v)\in S}|C^*(u,v)-C(u,v)|
&=&
\sup_{(u,v)\in S}|\expec_{u,v}[C(X/m,Y/m)-C(u,v)]|\\
&\leq&
\sup_{(u,v)\in S}\expec_{u,v}[|C(X/m,Y/m)-C(u,v)|]\\
&\leq&
\sup_{(u,v)\in S}\expec_{u,v}[|X/m-u|+|Y/m-v|]\\
&=&
(2/m)\sup_{u\in[0,1]}\expec_u[|X-mu|].
\end{eqnarray*}
In Lemma \ref{binomial} of the \hyperref[appendix]{Appendix}, we give the exact value of
$\sup_{u\in[0,1]}\expec_u[|X-mu|]$.
However,\vspace*{1pt} a~simple expression for an upper bound is given by H\"{o}lder's
inequality
\begin{eqnarray*}
\sup_{u\in[0,1]}2\expec_u[|X-mu|]/m
&\leq&
\sup_{u\in[0,1]}2\sqrt{\var_u[X]}/m\\
&=&
1/\sqrt{m}.
\end{eqnarray*}
\upqed\end{pf}

Bernstein copulas have appeared in the past literature and their
properties have
been extensively studied in Sancetta and Satchell \cite
{Sancetta1} and Sancetta and Satchell  \cite{Sanchetta2}. However,
in view of Lemma \ref{copulaapprox} and of the simplicity of indicator
functions, we subsequently use the indicator functions basis given
in~\eqref{c}
for $G$ in our model $C^*_P$ given by equation~\eqref{CP*}, and
$\mathcal{C}_m$ is the family of copulas generated by this model. Since
$P_{C_P^*}=P$ for any doubly stochastic matrix $P$, our model is rich
in the
sense that we have $\{P_{C}\dvt C\in\mathcal{C}\}=\{P_{C^*}\dvt C^*\in
\mathcal{C}_m\}$, which is the set of doubly stochastic matrices.

Notice that in a data reduction perspective, if $\{(u_k,v_k),
k=1,\ldots,n\}$ is
a sample from our model $C_P^*$, and if
$g=(g_1,\ldots,g_m)^{\top}$ represents the indicator functions
in~\eqref{c}, then $\{g(u_k)g(v_k)^{\top}, k=1,\ldots,n\}$ is a
sample from the $\operatorname{multinomial}(1, m^{-1}P)$ distribution. As in a~multinomial
experiment with probabilities given by $m^{-1}P$, the vector $(n_{ij})$
of cell
count statistics $ n_{ij}=\sum_{k=1}^n g_i(u_k)g_j(v_k),$
$i,j=1,\ldots,m, $ follows a $\operatorname{multinomial}(n, m^{-1}P)$ distribution
and is
sufficient for $P$.

The following lemma is used to define what we call Deheuvels'
estimator. The
estimator corresponds to the so-called \textit{bilinear extension} of
the empirical copula and has been considered by Deheuvels
\cite{Deheuvels2}, Nelsen \cite{Nelsen1}, Lemma 2.3.5, Genest and Ne{\v{s}}lehov{\'a} \cite{GN07}
and Ne{\v{s}}lehov{\'a} \cite{Nes07},
Section 5.

\begin{lemma}\label{Deheuvelsestimator}
Let $\{(x_i,y_i)\dvt i=1,\ldots,n\}$ be a sample, and let
$R=(r_{ij})$ be the
$n\times n$ matrix given by\vspace*{-3pt}
\[
r_{ij}=
(1/n)\sum_{k=1}^n \mathbf{1}\bigl(\operatorname{rank}(x_k)\leq i,
\operatorname{rank}(y_k)\leq j\bigr)\qquad\mbox{for }
i,j=1,\ldots,n.
\]
If we use the indicator basis~\eqref{c} with $m=n$ for $G$, then the copula
\[
\hat C_{\mathrm{DEH}}=n^2G^{\top} DRD^{\top} G\qquad \mbox{(Deheuvels' estimator)
}
\]
satisfies Deheuvels' constraint.\vspace*{-3pt}
\end{lemma}
%

\section{The prior distribution}\label{sec3}\vspace*{-3pt}

The choice of a prior concentrated on the approximation space
is delicate. The prior distribution is specified on $\mathcal{B}$, the
set of
doubly stochastic matrices of order $m$, $m>1$. Here, we adopt an objective
point of view and derive the Jeffreys prior. We also discuss two representations
of doubly stochastic matrices that can be useful for the specification
of other
prior distributions on~$\mathcal{B}$.

The set $\mathcal{B}$ is a convex polytope of dimension $(m-1)^2$. It
is known
in the literature as the Birkhoff polytope and has been the object of much
research. For instance, computing the exact value of its
volume is an outstanding problem in mathematics; it is known only for
$m\leq
10$ (see Beck and Pixton \cite{Beck1}).

The Fisher information matrix is obtained as follows. For $m>1$, let
$P\in\mathcal{B}$, and let $\mathcal{W}=(1/m)\mathcal{B}$. The
copula~\eqref{CP*} is a mixture of $m^2$ bivariate distribution functions
\begin{eqnarray*}
C_P^*(u,v)&=&mG(u)^{\top}PG(v)\\[-2pt]
&=&H(u)^{\top}WH(v)\\[-2pt]
&=&\sum_{i=1}^m\sum_{j=1}^mw_{ ij } H_i(u)H_j(v),
\end{eqnarray*}
where $W=(1/m)P\in\mathcal{W}$, and $H_i(u)=\int_0^uh_i(t)\,\mathrm{d}t$, for all
$u\in[0,1]$, with $h_i(\cdot)=mg_i(\cdot), i=1,\ldots,m$. The probability
density function
$c_P^*$ of the copula is thus
\begin{eqnarray*}
c_P^*(u,v)&=& \sum_{i=1}^m\sum_{j=1}^m w_{ij}h_i(u)h_j(v)\\
&=& 1+\sum_{i=1}^m\sum_{j=1}^m (w_{ij}-1/m^2)h_i(u)h_j(v)\\
&=& 1+\sum_{i=1}^{m-1}\sum_{j=1}^{m-1}
(w_{ij}-1/m^2)\bigl(h_i(u)-h_m(u)\bigr)\bigl(h_j(v)-h_m(v)\bigr).
\end{eqnarray*}
The last equality expresses the fact that there are $(m-1)^2 $ free parameters
in the model. Recall that we are considering the indicator functions
basis~\eqref{c} in our model. It follows that for all
$i_1,j_1,i_2,j_2=1,\ldots,m-1$,
\begin{eqnarray*}
\expec\biggl[\frac{-\partial^2\log c_P^*(u,v)}{\partial w_{i_1j_1}\,\partial
w_{i_2j_2}} \biggr]
&=&\int_0^1\int_0^1\frac{(h_{i_1}(u)h_{i_2}(u)+h_m^2(u))(h_{j_1}
(v)h_{j_2}(v)+h_m^2(v))}{c_P^*(u,v)}\,\mathrm{d}u\,\mathrm{d}v \\
&=& \cases{
1/w_{i_1j_1}+1/w_{i_1m}+1/w_{mj_1}+1/w_{mm}, &\quad if
$i_1=i_2,j_1=j_2$,\cr
1/w_{i_1m}+1/w_{mm}, &\quad if  $i_1=i_2,j_1\neq j_2$,\cr
1/w_{mj_1}+1/w_{mm}, &\quad if  $i_1\neq i_2, j_1=j_2$,\cr
1/w_{mm}, &\quad if  $i_1\neq i_2, j_1\neq j_2$.
}
\end{eqnarray*}
Although the information matrix is of order $(m-1)^2\times(m-1)^2$,
the following result shows how to reduce the computation of its determinant
to that of a matrix of order $(m-1) \times(m-1)$. The important reduction
provided by~\eqref{Fisher} is greatly appreciated when running an MCMC
algorithm,
which computes the determinant at every iteration. Most important, this
expression enables us to derive the main result of this paper, that is, Theorem
\ref{properJeffreys}. The proofs of these two results are quite
technical, so
we have put them in the \hyperref[appendix]{Appendix}.

\begin{lemma}\label{Jeffreysprior}
The Fisher information for $W=(w_{ij})_{i,j=1,\ldots,m}\in\mathcal
{W}$ is given
by
\begin{equation}\label{Fisher}
I(W)=\frac{\det((1/m)I-mV^{\top}V)}{m^m\det{(D_0)}\det{(D_1)}},
\end{equation}
where
\begin{eqnarray*}
V&=&(w_{ij})_{i=1,\ldots,m; j=1,\ldots,m-1},\\
D_0&=&\operatorname{diag}\bigl(w_{11},\ldots,w_{1(m-1)},\ldots,w_{(m-1)1},\ldots,w_{(m-1)(m-1)}\bigr)
\end{eqnarray*}
and
\[
D_1=\operatorname{diag}\bigl(w_{mm},w_{1m},\ldots,w_{(m-1)m},w_{m1},\ldots,w_{m(m-1)}\bigr).
\]
\end{lemma}

\begin{theorem}\label{properJeffreys}
The Jeffreys prior $\pi\varpropto I^{1/2}$ is proper.
\end{theorem}

Now, in order to specify different priors, we can consider the two following
representations.

\textit{The Hilbert space representation.}
Let $\mathcal{B}_0=\{P-(1/m)\mathbf{1}\mathbf{1}^{\top}\dvt
P\in\mathcal{B}\}$
and $\mathcal{V}=\operatorname{Span}{(\mathcal{B}_0)}$.
Consider the Frobenius inner product $\langle V_1,V_2\rangle=\operatorname{tr}(V_1V_2^{\top})$ on
$\mathcal{V}$.
Thus, $\mathcal{V}$ is an $(m-1)^2$-dimensional Hilbert space and an
orthonormal
basis is given by
$\{v_iv_j^{\top}\}_{i,j=1,\ldots,m-1},$ with
\[
v_i=\frac{1}{\sqrt{i(i+1)}}(\underbrace{1,\ldots,1}_i,-i,0,\ldots
,0)^{\top},\qquad
 i=1,\ldots,m-1.
\]
For every $P\in\mathcal{B}$, there exists a unique $(m-1)\times
(m-1)$ matrix
$A$ such that
\begin{equation}\label{HSrep}
P=m^{-1}\mathbf{1}\mathbf{1}^{\top}+G A G^{\top},
\end{equation}
where $G$ is the $m\times(m-1)$ matrix given by $G=(v_1,v_2,\ldots
,v_{m-1})$.
In this representation $A=G^{\top} P G$. Therefore, if we let
$\mathcal{B}'= G^{\top}\mathcal{B} G$, then we have a bijection
between $\mathcal{B}$
and~$\mathcal{B}'$. The set $\mathcal{B}'$ is a bounded convex subset of
$\mathbb{R}^{(m-1)^2}$ with positive Lebesgue measure. From this,
priors on
$\mathcal{B}$ can be induced by priors on $\mathcal{B}'$, and later
on, we
shall refer to the uniform prior on the polytope $\mathcal{B}$ as the uniform
distribution on $\mathcal{B}'$.
The above representation is also particularly useful to construct a Gibbs
sampler for distributions on the polytope.

\textit{The Birkhoff--von Neumann representation.}
Another decomposition is obtained by making use of the Birkhoff--von Neumann
theorem. Doubly stochastic matrices can be decomposed via convex
combinations of permutation matrices. In fact, $\mathcal{B}$ is the
convex hull
of the permutation matrices and these are precisely the extreme points (or
vertices) of $\mathcal{B}$. Furthermore, every $m\times m$ doubly stochastic
matrix $P$ is a convex combination of, at most, $k=(m-1)^2+1$ permutation
matrices; see Mirsky \cite{Mirsky1}. In other
words, if $\{\sigma_i\}
_{i=1}^{m!}$ is
the set of permutation matrices and if $P\in\mathcal{B}$, then there exists
$1\leq i_1 < \cdots< i_k \leq m!$ such that
$P=\sum_{j=1}^k\lambda_{i_j}\sigma_{i_j},$ for some weight vector
$(\lambda_{i_1},\ldots,\lambda_{i_k})$ lying in the
$(k-1)$-dimensional simplex
$\Lambda_k=\{(\lambda_1,\ldots,\lambda_k)\dvt0\leq\lambda_j, \mbox{
for all } j \mbox{ and }\sum_{j=1}^k\lambda_j=1\}$. A prior
distribution over
the polytope can be selected using a~discrete distribution over the set
$\{1\leq
i_1 < \cdots< i_k \leq m!\}$ and a continuous distribution over the simplex
$\Lambda_k$, such as a~Dirichlet distribution. See Melilli and Petris \cite{Melilli95} for
work in this direction.

\section{The MCMC algorithm}\label{sec4}

Let $\{(x_i,y_i), i=1,\ldots,n\}$ be a sample, where
each $(x_i,y_i)$ is a realization of the random couple $(X_i,Y_i)$,
$i=1,\ldots,n$, with dependence structure given by a copula $C$, and
with continuous marginal distributions $F_X$ and $F_Y$. If the marginal
distributions are known, then the transformed observations
$\tilde{x}_i=F_X(x_i)$ and $\tilde{y}_i=F_Y(y_i)$, $i=1,\ldots,n$,
are both
samples from a uniform distribution on $(0,1)$. If the marginal distributions
are unknown, then we follow Genest \textit{et al.} \cite{Genest2}
and consider the pseudo-observations
$\tilde{x}_i=(n/(n+1))\hat F_X(x_i)$ and $\tilde{y}_i=(n/(n+1))\hat
F_Y(y_i),{i=1,\ldots,n}$, where $\hat F_X$ and $\hat F_Y$ are the empirical
distributions. The algorithm below describes the transition kernel for the
Markov chain used to numerically evaluate the Bayesian estimator $\hat C$
associated to the Jeffreys prior $\pi$. The type of algorithm is
called Metropolis-within-Gibbs; see Gamerman and Lopes
\cite{Gamerman2006}. An
individual estimate is approximated by the sampling mean of the chain.

\begin{algorithm}[h!]
\textsf{Let
$T\geq1$ be the length of the chain, and at each iteration $t$, $1\leq
t \leq
T$, let $P_t$ be the current doubly stochastic matrix. From
representation~\eqref{HSrep} in the previous section with
$A=(a_{kl})_{k,l=1,\ldots,m-1}$,
\[
P_t-(1/m)\mathbf{1}\mathbf{1}^{\top}=\sum_{k=1}^{m-1}\sum_{l=1}^{m-1}a_{kl}
v_kv_l^{\top}.
\]
Repeat for $i,j=1,\ldots,m-1$:
\begin{enumerate}[1.]
\item Select direction $v_iv_j^{\top}$ and compute the interval
$I_{ij}\subset
\mathbb{R}$ as follows:
\begin{enumerate}[1.1]
\item[1.1] For every $p,q=1,\ldots,m$, find the largest interval
$I_{ij}^{(p,q)}$ such that
\[
\epsilon_{ij}v_i^{(p)}v_j^{(q)}\geq-1/m-\sum_{k=1}^{m-1}\sum_{l=1}^{m-1}a_{
kl}v_k^{(p)}v_l^{(q)}\qquad \mbox{for all } \epsilon_{ij}\in I_{ij}^{(p,q)}.
\]
\item[1.2] Take $\Gamma_{ij}=\bigcap_{p,q}\Gamma_{ij}^{(p,q)}.$
\end{enumerate}
\item Draw $\epsilon_{ij}$ from the uniform distribution on $I_{ij},$
and set
$a_{ij}'=a_{ij}+\epsilon_{ij}$ and $a_{kl}'=a_{kl}$, for every $k
\neq i, l \neq j$. The proposed doubly stochastic matrix is given by
\[
P_t^{\mathrm{prop}}=(1/m)\mathbf{1}\mathbf{1}^{\top}+\sum
_{k=1}^{m-1}\sum_{l=1}^{m-1}
a_{kl}'v_kv_l^{\top}.
\]
\item Accept $P_{t+1}=P_t^{\mathrm{prop}}$ with probability
\begin{equation} \label{alpha}
\alpha(P_t,P_t^{\mathrm{prop}})=\min\biggl\{1,\frac{\pi(P_t^{\mathrm{prop}})L(P_t^{
\mathrm{prop}}\mid\tilde{x},\tilde{y})}{\pi(P_t)L(P_t\mid\tilde
{x},\tilde{y})}\biggr\},
\end{equation}
where $L(\cdot\mid\tilde{x},\tilde{y})$ is the likelihood derived from
expression~\eqref{CP*}.
\end{enumerate}}
\end{algorithm}

\begin{figure}
\centering
\begin{tabular}{@{}cc@{}}

\includegraphics{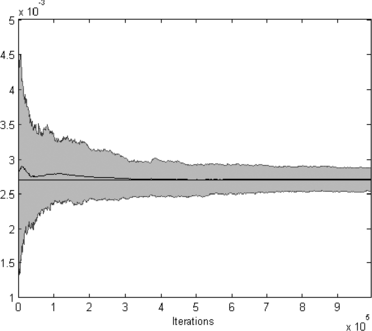}
&\includegraphics{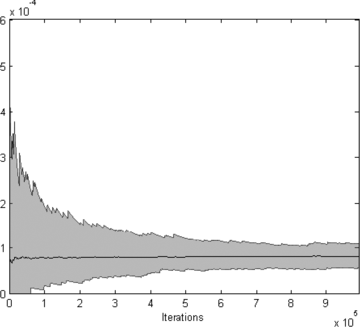}\\
(a)& (b)
\end{tabular}
\caption{Convergence of 1000 parallel MCMC runs for the
probability of the largest ball contained in the polytope $\mathcal{B}$
with $m=4$. Shaded region represents the range of the entire set of
approximations at each iteration. Figure (a) is the
convergence for the
probability in the case of the uniform distribution. The flat line, in
this case, corresponds to the true probability $p\approx0.0027$.
Figure (b) is the
same for the Jeffreys prior.}\label{Fig1}
\end{figure}

Note that the above algorithm could also be used with any prior
specified via
the Hilbert space representation described in the previous section, including
the uniform prior on the polytope $\mathcal{B}$ described in the
previous section. In particular, it could be
adapted to draw random doubly stochastic matrices according to such priors
by replacing the acceptance probability~\eqref{alpha} with
\[
\alpha(P_t,P_t^{\mathrm{prop}})=\min\biggl\{1,\frac{\pi(P_t^{\mathrm{prop}})}{
\pi(P_t)}\biggr\}.
\]
In order to further describe the Jeffreys prior, we use the algorithm
to approximate the probability of the largest ball contained in $\mathcal
B$ with respect to the Euclidean distance on~$\mathcal B$. This distance
may be computed using the Frobenius inner product described in the
previous section. The largest
ball has radius $1/(m-1)$, where $m>1$ is the size of the doubly stochastic
matrix. Although this probability can be obtained exactly for the
uniform distribution, we nevertheless approximate it using our algorithm,
meanwhile providing some validation of the MCMC algorithm. Figure
\ref{Fig1} shows the results we get for $m=4$.

%

\begin{figure}
\centering
\begin{tabular}{@{}cc@{}}

\includegraphics{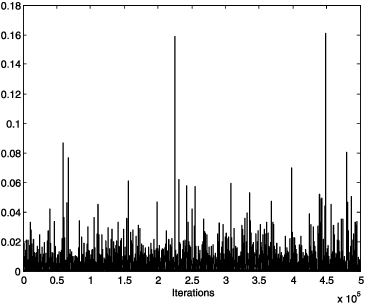}
 & \includegraphics{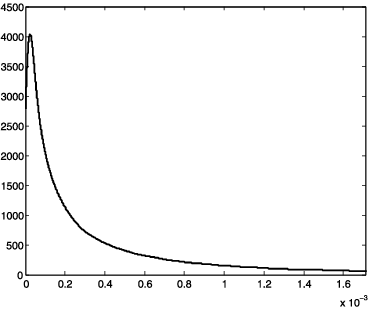}\\
(a)& (b)\\ [6pt]

\includegraphics{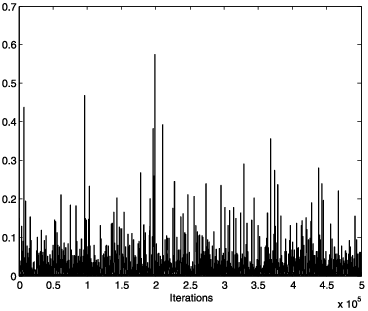}
 & \includegraphics{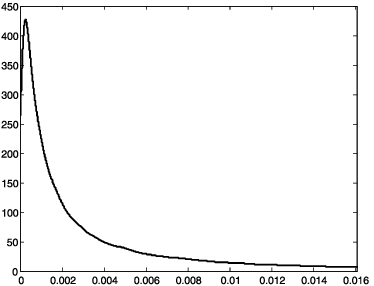}\\
(c) & (d)
\end{tabular}
\caption{Plots of samples and density estimates of the radius
(Euclidean distance of the doubly stochastic matrix from the center of
the polytope
$\mathcal{B}$), on the interval $[0,q_{95}]$, where $q_{95}$ is the 95th
quantile of its distribution. Figures (a) and (b) are results when sampling from the uniform
prior and figures (c) and (d) are
those of the Jeffreys prior. Here $m=4$.}\label{Fig2}
\end{figure}

Notice that this probability is much smaller for the Jeffreys prior, because
it distributes more mass towards the extremities of the polytope than the
uniform prior does. This may also be observed by plotting the density estimates
of the radius of the doubly stochastic matrix, that is, the Euclidean distance
of the doubly stochastic matrix from the center of the polytope $\mathcal{B}$.
These are shown in Figure \ref{Fig2}.

%
%

\section{Simulation experiments}\label{sec5}

The goal of the experiment is to study the performance of our
estimator on artificial data sets generated from various bivariate
distributions. We provide evidence that the estimators derived from our model
give good results in general, and most important, that the Jeffreys
prior is a~reasonable choice.

Six parametric families of copulas are considered:
\begin{longlist}
\item[1.]\textit{Clayton family}: $C_{\theta}(u,v)=\{\max(0,u^{-\theta
}+v^{-\theta}-1)\}^{-1/\theta},\ \theta\geq-1, \theta\neq0,$
\item[2.]\textit{Gumbel family}: $C_{\theta}(u,v)=\exp[-\{(-\log
u)^{\theta}+(-\log v)^{\theta}\}^{1/\theta}],\ \theta\geq1,$\vspace*{2pt}
\item[3.]\textit{Frank family}:
$C_{\theta}(u,v)=-\frac{1}{\theta}\log\{1+\frac{(\mathrm{e}^{-\theta
u}-1)(\mathrm{e}^{-\theta v}-1)}{\mathrm{e}^{-\theta}-1}\},\ \theta\neq0,$\vspace*{2pt}
\item[4.]\textit{Gaussian family}: $C_{\theta}(u,v)=\Phi_{\theta}(\Phi
^{-1}(u),\Phi^{-1}(v)),\ |\theta| \leq1,$ where $\Phi_{\theta
}$ is the standard bivariate Gaussian cumulative distribution function
with correlation coefficient $\theta$, and~$\Phi^{-1}$ is the inverse
of the univariate standard normal cumulative distribution function.
\item[5.]\textit{Gaussian cross family}: $C_{\theta}^{\times
}(u,v)=1/2(C_{\theta}(u,v)-C_{\theta}(u,1-v)+u),\ |\theta| \leq
1$, where~$C_{\theta}$ belongs to the Gaussian family.
\item[6.]\textit{Gaussian diamond family}:
\[
C_{\theta}^{\diamond}(u,v)=\cases{
C_{\theta}^{\times}(u+1/2,v)-C_{\theta}^{\times}(1/2,v), &\quad
if $u\leq1/2$,\cr
C_{\theta}^{\times}(u-1/2,v)+v-C_{\theta}^{\times}(1/2,v), &\quad
if $u>1/2$,}
\]
where $|\theta| \leq1$ and $C_{\theta}^{\times}$ belongs to the
Gaussian cross family.
\end{longlist}

\begin{figure}
\centering
\begin{tabular}{@{}cc@{}}

\includegraphics{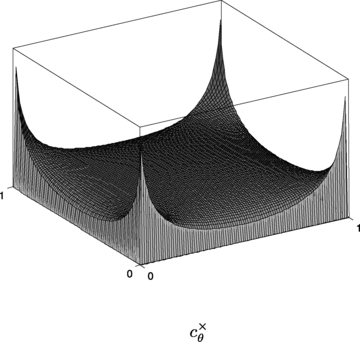}
 & \includegraphics{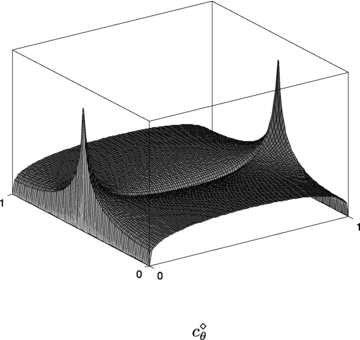}\\[3pt]
(a)& (b)
\end{tabular}
\caption{Densities of the Gaussian cross copula and the Gaussian
diamond copula with $\theta=0.5$.}\label{Fig3}
\vspace*{3pt}
\end{figure}

For the Clayton, Frank and Gaussian families, values of the parameter
away from 0 indicate departure from independence, while a parameter
away from 1 indicates departure from independence for the Gumbel
family. These four families are among the popular ones in the
literature, but the last two families are not, and so we now describe
them in more detail. The Gaussian cross family is obtained by the
following: Let $(U,V)$ be a~random vector with uniform margins and
with the Gaussian copula $C_{\theta}$ as its joint distribution.
Let~$W$ be an independent uniformly
distributed random variable, and consider the random vector
\[
(U_{\theta}^{\times},V_{\theta}^{\times})=(U,V)\mathbf{1}(W\leq
1/2)+(U,1-V)\mathbf{1}(W>1/2).
\]
The
distribution of $(U_{\theta}^{\times},V_{\theta}^{\times})$ is
given by the Gaussian cross copula. Here, the superscript $\times$ is
to highlight the ``cross-like'' dependence structure; see Figure \ref{Fig3}(a)
for a plot of its density when $\theta
=0.5$. The Gaussian diamond family corresponds to the distributions of
the random vectors
\[
(U_{\theta}^{\diamond},V_{\theta}^{\diamond})=\bigl(U_{\theta}^{\times
}+1/2(\mbox{mod } 1),V_{\theta}^{\times}\bigr)\qquad\mbox{for each
} |\theta| \leq1.
\]
See Figure \ref{Fig3}(b) for an illustration of its
density when $\theta=0.5$.

An extensive simulation experiment is carried out in two parts. In the first
part, we consider the case of known marginal distributions and use bivariate
data sampled from the copula families above. For each family, we
consider 11 models corresponding to equally spaced parameter values in
some interval. For the first four families, the interval is determined
so that the Kendall's $\tau$ values associated to the particular
models range between~0 and $2/3$; see the simulation in Silva and Lopes \cite{Silva1}.
Kendall's $\tau$ associated with a~copula~$C$ is the dependence
measure defined by
\[
\tau=4\int_0^1\int_0^1C(u,v)\,\mathrm{d}C(u,v)-1.
\]
The values of Kendall's $\tau$ for the first four families are
respectively given by \mbox{$\tau=\theta/(\theta+2)$}, $\tau=1-1/\theta$,
$\tau=1-(4/\theta)[1-D_1(\theta)]$, where $D_1$ is the Debye
function and $\tau=(2/\pi)\arcsin(\theta)$. For families 5 and 6,
we consider the models corresponding to 11 values of $\theta$ ranging
between 0 and 1. In the second part of the experiment, we simulate an
unknown margins situation. We focus on families 4, 5 and 6 and
consider 11 equally spaced values of $\theta$ ranging between 0 and 1
for the copula models. Here, a Student $t$ with seven degrees of
freedom and a chi-square with
four degrees of freedom are considered as the first and second margins,
respectively.

\begin{figure}[t!]
\centering
\begin{tabular}{@{}cc@{}}

\includegraphics{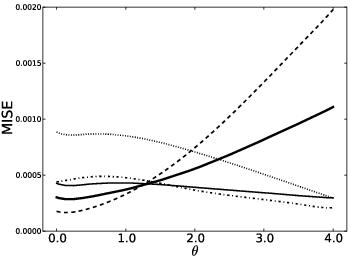}
 & \includegraphics{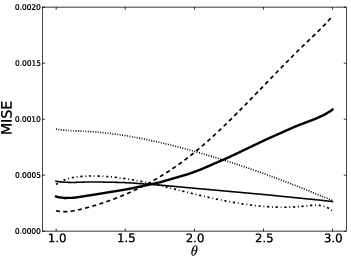}\\
(a) Family 1 & (b) Family 2\\ [6pt]

\includegraphics{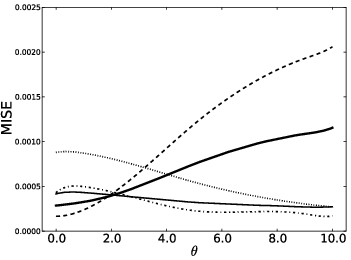}
 & \includegraphics{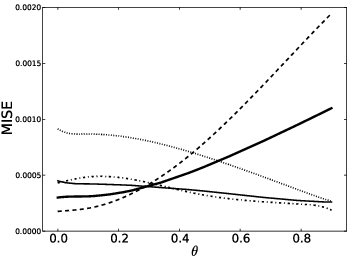}\\
(c) Family 3 & (d) Family 4\\ [6pt]

\includegraphics{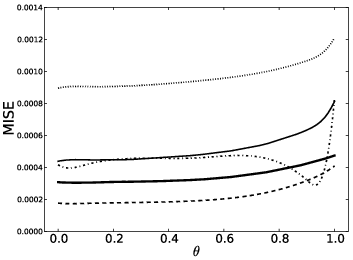}
 & \includegraphics{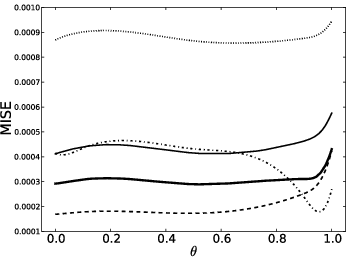}\\
(e) Family 5 & (f) Family 6
\end{tabular}
\caption{Plots of MISE against $\theta$ in the known margins case.
The MISE is approximated using 1000 samples of size $n=30$. Thick
solid line is the MISE of the Bayes estimator using the Jeffreys prior,
dashed line is that of the Bayes estimator using the uniform prior,
dashed--dotted line is the MISE of the MLE, while dotted and thin solid
line is the MISE of Deheuvels' and the Gaussian kernel estimators,
respectively. }\label{Fig4}
\end{figure}

\begin{figure}
\centering
\begin{tabular}{@{}cc@{}}

\includegraphics{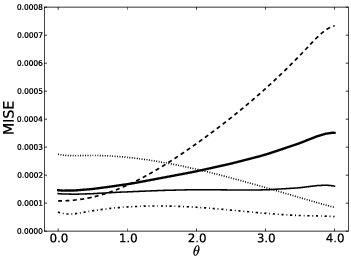}
 & \includegraphics{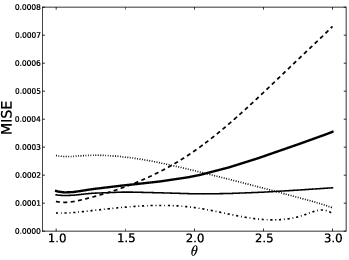}\\
(a) Family 1 & (b) Family 2\\ [6pt]

\includegraphics{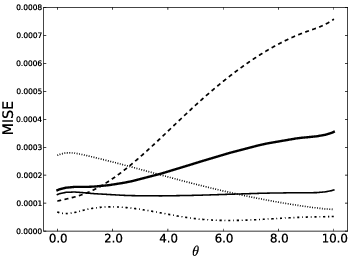}
 & \includegraphics{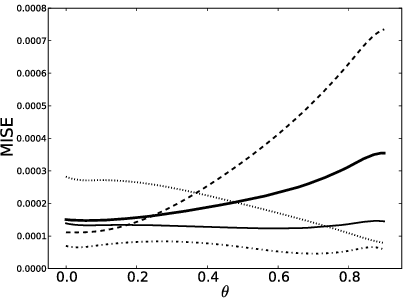}\\
(c) Family 3 & (d) Family 4\\ [6pt]

\includegraphics{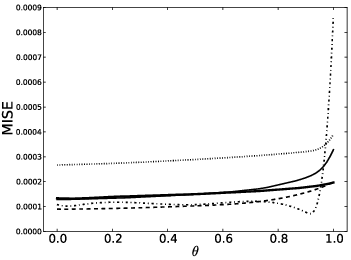}
 & \includegraphics{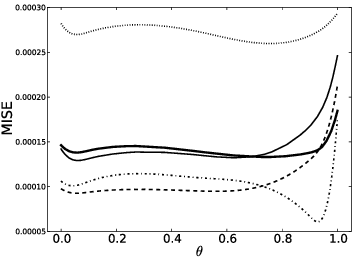}\\
(e) Family 5 & (f) Family 6
\end{tabular}
\caption{Plots of MISE against $\theta$ in the known margins case.
The MISE is approximated using 1000 samples of size $n=100$. Thick
solid line is the MISE of the Bayes estimator using the Jeffreys prior,
dashed line is that of the Bayes estimator using the uniform prior,
dashed--dotted line is the MISE of the MLE, while dotted and thin solid
line is the MISE of Deheuvels' and the Gaussian kernel estimator,
respectively.}\label{Fig5}
\end{figure}

\begin{figure}[t!]
\centering
\begin{tabular}{@{}cc@{}}

\includegraphics{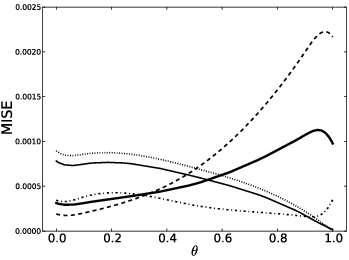}
 & \includegraphics{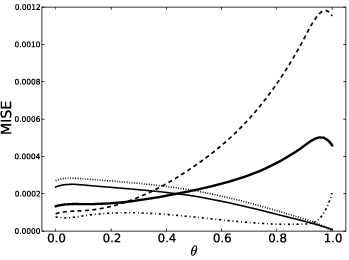}\\
(a) Model 4, $n=30$ & (b) Model 4, $n=100$\\ [6pt]

\includegraphics{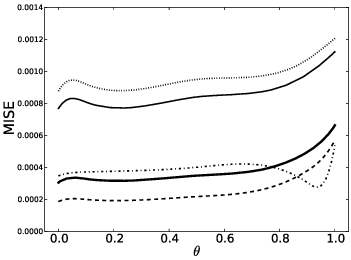}
 & \includegraphics{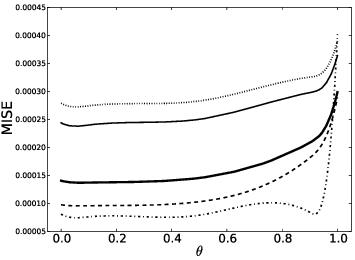}\\
(c) Model 5, $n=30$ & (d) Model 5, $n=100$\\ [6pt]

\includegraphics{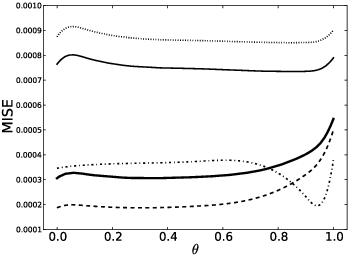}
 & \includegraphics{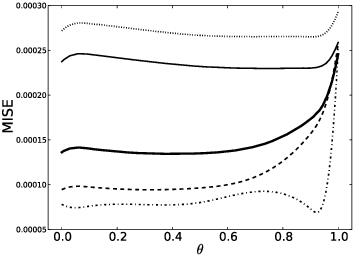}\\
(e) Model 6, $n=30$ & (f) Model 6, $n=100$
\end{tabular}
\caption{Plots of MISE against $\theta$ in the unknown margins case.
The MISE is approximated using 1000 samples each of sizes $n=30$
and $n=100$. Thick solid line is the MISE of the Bayes estimator using
the Jeffreys prior, dashed line is that of the Bayes estimator using
the uniform prior, dashed--dotted line is the MISE of the MLE, while
dotted and thin solid line is the MISE of Deheuvels' and the Gaussian
kernel estimators, respectively.}\label{Fig6}
\end{figure}

In the experiment, 1000 samples of both sizes $n=30$ and $n=100$
are generated from each model. For every data set, the copula function
is estimated using five estimators. The first two are the Bayes
estimators associated to the Jeffreys and the uniform priors,
respectively. For the uniform prior, we mean the uniform distribution
on $\mathcal{B}'$ defined in Section \ref{sec3}, we use the bijection $\mathcal
{B}=m^{-1}\mathbf{1}\mathbf{1}^{\top}+G\mathcal{B}'G^{\top}$ given
by expression~\eqref{HSrep}. The third estimator is the maximum
likelihood estimator (MLE) from our model $C_{\hat P}^*$, where~$\hat
{P}$ maximizes the likelihood derived from expression~\eqref{CP*}.
This estimator is evaluated numerically. For the above three
estimators, we take $m=6$ as the order of the doubly stochastic matrix
in our model. Finally, we consider the two frequentist estimators, that
is, Deheuvels' estimator given in Theorem \ref{Deheuvelsestimator} and
the Gaussian kernel estimator described in the \hyperref[intro]{Introduction}. Values of
the bandwidth for the latter estimator are based on the commonly used
rule of thumb: $h= s_i n^{-1/5}$, where $s_i$, $i=1,2$, is the sample
standard deviation of the $i$th margin; see Fermanian and Scaillet \cite{Fermanian1} and
Sheather \cite{Sheather04}. Figures~\ref{Fig4},~\ref{Fig5} and \ref{Fig6}
report the values of the mean integrated squared errors,
\[
\operatorname{MISE}(\widehat{C})=\expec
\biggl[\int_0^1\int_0^1\bigl(\widehat{C}(u,v)-C(u,v)\bigr)^2\,\mathrm{d}u\,\mathrm{d}v \biggr],
\]
for the five estimators as a function of the parameter $\theta$.

%

%

%

As the results indicate, the Bayesian approach outperforms Deheuvels'
estimator and the kernel estimator near independence for the Clayton,
Gumbel, Frank and Gaussian families. Unfortunately, this is not
necessarily the case when the value of the parameter increases, that
is, when the true copula approaches the Fr\'{e}chet--Hoeffding upper
bound, also called the comonotone copula, corresponding to (almost
sure) perfect positive linear dependence. For families 5 and 6, the
Bayes estimators outperform both frequentist estimators when the sample
size is small ($n=30$). One remarkable feature that appears when
comparing the results obtained in the known margins case with the
results obtained in the unknown margins case is the decrease in
performance of the kernel estimator. Recall that the latter estimator
is the only one for which the invariance property mentioned in the
Introduction does not hold. The other estimators seem to behave
similarly when comparing their resulting MISE in the known margins case
with their MISE in the unknown margins case. Notice the resemblance in
shape of the MISE for Deheuvels' estimator and the kernel\vadjust{\goodbreak} estimator in
the unknown margins
cases. Finally, the performance of the MLE is worth mentioning, since
in many
cases it has the smallest MISE, especially for large values of $\theta$.
This is because the MLE will go on the boundary of the parameter
space easily,
while the Bayes estimator will always stay away from the boundary with
the types
of priors that we have selected. However, if such an extreme case is to happen
in a real life problem, it is probable that the practitioner has some
insight on
the phenomenon beforehand, and may choose to work with a more appropriate
(subjective) prior.\vspace*{-3pt}

\section{Discussion}\label{sec6}\vspace*{-3pt}

Two points need to be further discussed. First, our methodology is purely
Bayesian only when the marginal distributions are known. When these are
unknown, our methodology is empirical Bayes. In fact, in this case we
propose a
two-step procedure by first estimating the marginsvia the empirical marginal
distributions and  then plugging them in as the true distributions thereafter.
We have chosen to do this because it is common practice to do so (see
Genest \textit{et al.} \cite{Genest2}), it is simple to implement,
it is robust against
outliers and our estimator
is consequently invariant under increasing transformations of the margins.
One way to propose a purely Bayesian estimator by using our model for the
copula is to use finite mixtures for the margins. This way, if the densities
used in the latter mixtures have disjoint supports, then the Jeffreys
prior for
the mixing weights has a simple form and is proper; see
Bernardo and Gir\'{o}n \cite{Bernardo88}. Now
by selecting independent Jeffreys priors for the margins and for the
copula, the resulting prior is proper as well.

Finally, our models given by the approximation spaces $\mathcal{C}_m$,
$m>1$, are
called \textit{sieves} by some authors; see Grenander
\cite{Grenander81}. In
the present paper,
we have chosen to work with a fixed sieve, so this makes our model
finite dimensional. In this case, the methodology falls in the \textit
{semi-parametric approach} described in the \hyperref[intro]{Introduction}. Here, the
rather subjective choice of the sieve to work with can be viewed as a
weakness of the proposed methodology. On the other hand, by using the
entire set of sieves, we can construct a nonparametric model for the
copula that can, in some
sense, respect the infinite-dimensional nature of the copula functions. In
fact, if we take $\mathcal{C}=\bigcup_{m>1}\mathcal{C}_m$, then $\mathcal{C}$ is
dense in
the space of copulas. Our Bayesian methodology can be easily adapted here.
This can be achieved by selecting an infinite support prior for the
model index
$m$ and using our methodology inside each model. The Bayesian estimator
becomes an infinite mixture of the estimators proposed in this
paper (one for each model $m$), where the mixing weights are given by the
posterior probabilities of the models.\vspace*{-3pt}

\begin{appendix}\label{appendix}
\renewcommand{\theequation}{\arabic{equation}}
\setcounter{equation}{10}
\section*{Appendix}\vspace*{-3pt}

\begin{pf*}{Proof of Lemma \ref{Jeffreysprior}}
Here we show how to compute $I(W)=\det
(\expec[\frac{-\partial^2\log c_P^*(u,v)}{\partial W^2}
])$
efficiently. First, notice that $A=\expec[\frac{-\partial^2\log
c_P^*(u,v)}{\partial W^2}]$ can be written as
$A=D_0^{-1}+CD_1^{-1}C^{\top},$
where
\begin{eqnarray*}
D_0&=&\operatorname{diag}\bigl(w_{11},\ldots,w_{1(m-1)},\ldots,w_{(m-1)1},\ldots,w_{
(m-1)(m-1)}\bigr),\\
D_1&=&\operatorname{diag}\bigl(w_{mm},w_{1m},\ldots,w_{(m-1)m},w_{m1},\ldots,w_{m(m-1)}\bigr)\vadjust{\goodbreak}
\end{eqnarray*}
and
\[
C_{(m-1)^2\times(2m-1)}=\pmatrix{
\mathbf{1}_{m-1} & \mathbf{1}_{m-1} & 0\mathbf{1}_{m-1} & \cdots&
I_{m-1} \cr
\mathbf{1}_{m-1} & 0\mathbf{1}_{m-1} & \mathbf{1}_{m-1} & \cdots&
I_{m-1}\cr
\mathbf{1}_{m-1} & 0\mathbf{1}_{m-1} & 0\mathbf{1}_{m-1} & \cdots&
I_{m-1} \cr
\vdots& \vdots& \vdots& \vdots& \vdots\cr
\mathbf{1}_{m-1} & 0\mathbf{1}_{m-1} & 0\mathbf{1}_{m-1} & \cdots&
I_{m-1} \cr
}.
\]
Thus, $\det A=\det(D_1+C^{\top}D_0C)/(\det D_0 \det D_1)$. If we let
$B=(w_{ij})_{i,j=1,\ldots,m-1}$, then since $\sum_{i=1}^mw_{ij}=1/m$,
for all
$j=1,\ldots,m$, and $\sum_{j=1}^mw_{ij}=1/m$ for all $i=1,\ldots,m$,
\[
D_1+C^{\top}D_0C=\pmatrix{
1-2(1/m-w_{mm}) & \mathbf{1}^{\top}B^{\top} & \mathbf{1}^{\top}B \cr
B\mathbf{1} & (1/m)I & B \cr
B^{\top}\mathbf{1} & B^{\top} &(1/m)I \cr
}.
\]
By elementary row and column operations, we get
\[
\det(D_1+C^{\top}D_0C)=\det\pmatrix{
1 & (1/m)\mathbf{1}^{\top} & (1/m)\mathbf{1}^{\top} \cr
(1/m)\mathbf{1}& (1/m)I & B \cr
(1/m)\mathbf{1}& B^{\top} & (1/m)I \cr
},
\]
so that
\begin{eqnarray*}
&&\det(D_1+C^{\top}D_0C)\\
&&\quad= \det\left( \pmatrix{
(1/m)I & B \cr
B^{\top} & (1/m)I
} -
(1/m^2)\mathbf{1}_{2(m-1)}\mathbf{1}_{2(m-1)}^{\top}
\right)\\
&&\quad= \det\pmatrix{
(1/m)\bigl(I-(1/m)\mathbf{1}\mathbf{1}^{\top}\bigr) & B
-(1/m^2)\mathbf{1}\mathbf{1}^{\top}\cr
B^{\top}-(1/m^2)\mathbf{1}\mathbf{1}^{\top} &
(1/m)\bigl(I-(1/m)\mathbf{1}\mathbf{1}^{\top}\bigr)}\\
&&\quad=\det
\bigl((1/m)\bigl(I-(1/m)\mathbf{1}\mathbf{1}^{\top}\bigr)\bigr)\det
\bigl(\bigl[(1/m)\bigl(I-(1/m)\mathbf{1}
\mathbf{1}^{\top}\bigr)\bigr]\\
&&\qquad\phantom{\det
\bigl(}{}-[B^{\top}-(1/m^2)\mathbf{1}\mathbf{1}^{\top
}]\bigl[(1/m)\bigl(I-(1/m)\mathbf{1}
\mathbf{1}^{\top}\bigr)\bigr]^{-1}[B -(1/m^2)\mathbf{1}\mathbf{1}^{\top
}]\bigr).
\end{eqnarray*}
Finally, $\det((1/m)(I-(1/m)\mathbf{1}\mathbf{1}^{\top})
)=(1/m)^m \mbox{ and
} [(1/m)(I-(1/m)\mathbf{1}\mathbf{1}^{\top})]^{-1}=m(I+\mathbf
{1}\mathbf{1}^{\top})$, thus
\[
\det(D_1+C^{\top}D_0C)=(1/m)^m\det\bigl((1/m)I-mV^{\top}V\bigr),
\]
%
where $V=(w_{ij})_{i=1,\ldots,m; j=1,\ldots,m-1}$.
\end{pf*}

\begin{pf*}{Proof of Theorem \ref{properJeffreys}}
We prove that the Jeffreys prior is proper. Consider the following
partition of
$V$, $V=(V_1 V_2\cdots V_{m-1})$, where each $V_j$ is a vector,
$j=1,\ldots,m-1$. The matrix $(1/m)I-mV^{\top}V$ is symmetric,
non-negative and
semi-definite, so that by Hadamard's inequality, we have
\begin{eqnarray*}
\det\bigl((1/m)I-mV^{\top}V\bigr)&\leq& \prod_{j=1}^{m-1}(1/m-m\|V_j\|^2)\\
&=& \prod_{j=1}^{m-1} \biggl(2m \sum_{1\leq i<k\leq m}w_{ij}w_{kj} \biggr)\\
&=& (2m)^{m-1} \sum_{1\leq i_{m-1}<k_{m-1}\leq m}\cdots\sum_{1\leq
i_{1}<k_{1}\leq m}\prod_{j=1}^{m-1}w_{i_jj}w_{k_jj}.
\end{eqnarray*}
For any $W\in\mathcal{W}$, we have
\begin{eqnarray*}
\sqrt{I(W)}
&=&\sqrt{\det\bigl((1/m)I-mV^{\top}V\bigr)}\Big/\sqrt{m^m\prod_{i,j=1}^m
w_{ij}}\\
&\leq&
\sqrt{2^{m-1}/m}
\Biggl\{
\sum_{1\leq i_{m-1}<k_{m-1}\leq m}\cdots\sum_{1\leq i_{1}<k_{1}\leq
m}\prod_{j=1}^{m-1}w_{i_jj}w_{k_jj}\Biggr\}^{1/2}
\bigg/\sqrt{\prod_{i,j=1}^m w_{ij}}\\
&\leq&
\sqrt{2^{m-1}/m}
\sum_{1\leq i_{m-1}<k_{m-1}\leq m}\cdots\sum_{1\leq i_{1}<k_{1}\leq m}
\Biggl\{\prod_{j=1}^{m-1}w_{i_jj}w_{k_jj}
\Biggr\}^{1/2}
\bigg/\sqrt{\prod_{i,j=1}^m w_{ij}}\\
&=&
\sqrt{2^{m-1}/m}
\sum_{\alpha\in\mathcal{A}}
\prod_{i,j=1}^m w_{ij}^{\alpha_{ij}-1},
\end{eqnarray*}
where $\mathcal{A}=
\{\alpha_{ij}\in\{1/2,1\}\dvt i,j=1,\ldots,m,\alpha_{+j}=m/2+1,
j=1,\ldots
,m-1 \mbox{ and } \alpha_{+m}=m/2\},$ with
$\alpha_{+j}=\sum_{i=1}^m\alpha_{ij}$, for all $j=1,\ldots,m$. We
need to show
that the integral of $\prod_{i,j=1}^m w_{ij}^{\alpha_{ij}-1}$ is
finite for all $\alpha\in\mathcal{A}$. The integration is made with
respect to
$w_{ij}, i\vee j <m$, the free variables. For any permutation
matrices $P_1$
and $P_2$, the transformation $W \mapsto P_1 W P_2$ is a one-to-one
transformation from $\mathcal W$ onto $\mathcal W$, and the Jacobian, in
absolute value, is equal to one. Therefore, it is sufficient to verify
that the
integral of $\prod_{i,j=1}^m w_{ij}^{\alpha_{ij}-1}$ is finite for all
$\alpha\in\mathcal{A}_0$, where
$\mathcal{A}_0=\{\alpha\in\mathcal{A}\dvt\alpha_{m-1m}=\alpha
_{mm}=1\}.$
The idea is to decompose the multiple integral into $m-2$ iterated integrals
over the sections given by
\[
\mathcal{W}_k=\{w_{ij}\geq0\dvt i\wedge j=k,i\vee j\leq m,
w_{k+}=w_{+k}=1/m\},\qquad k=1,\ldots, m-2,
\]
and
\[
\mathcal{W}_{m-1}=\{w_{ij}\geq0\dvt
i,j=m-1,m,
w_{m-1 +}=w_{+ m-1}=w_{m +}=w_{+ m}=1/m\}.
\]
Here, the set $\mathcal{W}_1$ is fixed, the sets
$\mathcal{W}_k$ are parameterized by
$\{w_{ij}\geq0\dvt i\wedge j<k, i\vee j=k\}$,
$k=2,\ldots,m-2$, and $\mathcal{W}_{m-1}$
is parameterized by
$\{w_{ij}\geq0\dvt i\wedge j<m-1, i\vee j=m-1,m\}$.
By Fubini's theorem, for any non-negative function $f$, we can write
\[
\int_\mathcal{W} f(W)\prod_{i,j=1}^{m-1}\mathrm{d}w_{ij}=
\int_\mathcal{W_1}\biggl\{
\cdots
\int_\mathcal{W_{m-1}}\{
f(W)\,\mathrm{d}w_{m-1 m-1}\}
\cdots\biggr\}
\prod_{i\wedge j=1,i\vee j<m}\mathrm{d}w_{ij}.
\]

The next step consists in finding finite functions $c_k$, $k=1,\ldots
,m-1$, on
$\mathcal{A}_0$, such that
\[
\int_\mathcal{W_k}
\prod_{i\wedge j=k,i\vee j\leq m} w_{ij}^{\alpha_{ij}-1}
\prod_{i\wedge j=k,i\vee j<m}\mathrm{d}w_{ij}
\leq c_k(\alpha),
\]
for all $\alpha\in\mathcal{A}_0$, uniformly on
$\{w_{ij}\geq0\dvt i\wedge j<k, i\vee j=k\}$,
for $k=1,\ldots,m-2$,
and uniformly on
$\{w_{ij}\,{\geq}\,0\dvt i\,{\wedge}\,j\,{<}\,m\,{-}\,1, i\,{\vee}\,j\,{=}\,m\,{-}\,1,m\}$,
for $k\,{=}\,m\,{-}\,1$.
This will give us~that
\[
\int_\mathcal{W}
\prod_{i,j\leq m} w_{ij}^{\alpha_{ij}-1}
\prod_{i,j<m} \mathrm{d}w_{ij}
\leq\prod_{k=1}^{m-1}c_k(\alpha),
\]
for all $\alpha\in\mathcal{A}_0$.

Let $a=0\vee\{\sum_{\ell<m-1}(w_{\ell m}-w_{m-1\ell})\}\vee
\{\sum_{\ell<m-1}(w_{m \ell}-w_{\ell m-1})\}$ and
$b=1/m-\{(\sum_{\ell<m-1}w_{\ell m-1})\vee(\sum_{\ell
<m-1}w_{m-1\ell})\}$.
If $a>b$, the set $\mathcal W_{m-1}$ is empty. Suppose that~$\mathcal W_{m-1}$
is not empty and
$\alpha\in\mathcal A_0$. Let $b_0=1/m-\sum_{\ell<m-1}w_{\ell
m-1}$. We have
\begin{eqnarray*}
\int_\mathcal{W_{m-1}}
\prod_{i,j=m-1, m} w_{ij}^{\alpha_{ij}-1}\,\mathrm{d}w_{m-1 m-1}
&=&
\int_a^b u^{\alpha_{m-1 m-1}-1}(b_0-u)^{\alpha_{m m-1}-1}\,\mathrm{d}u\\[-2pt]
&\leq&
\int_0^{b_0} u^{\alpha_{m-1 m-1}-1}(b_0-u)^{\alpha_{m m-1}-1}\,\mathrm{d}u\\[-2pt]
&=&b_0^{\alpha_{m-1 m-1}+\alpha_{m m-1}-1}B(\alpha_{m-1
m-1},\alpha_{m
m-1})\\[-2pt]
&\leq&
B(\alpha_{m-1 m-1},\alpha_{m m-1})\\[-2pt]
&=&
c_{m-1}(\alpha).
\end{eqnarray*}
For $k=1,\ldots,m-2$ and $\alpha\in\mathcal A_0$ we can take
\[
c_k(\alpha)=
\Biggl(B\Biggl(\alpha_{kk},\sum_{i=k+1}^m\alpha_{ik}\Biggr)
+B\Biggl(\alpha_{kk},\sum_{j=k+1}^m\alpha_{kj}\Biggr)\Biggr)
\frac{\prod_{i>k}\Gamma(\alpha_{ik})
\prod_{j>k}\Gamma(\alpha_{kj})}
{\Gamma(\sum_{i>k}\alpha_{ik})\Gamma(\sum_{j>k}\alpha_{kj}
)}.
\]
The justification is given by Lemma \ref{Dirichlet}.
\end{pf*}

\begin{lemma}
\label{Dirichlet}
If $0<a,b\leq1, m\geq3, \alpha>0$,
\begin{eqnarray*}
\beta_j&>&0,\qquad j=1,\ldots,m-1,\qquad
\mbox{with } \beta=\sum_{j=1}^{m-1}\beta_j\geq1,\\[-2pt]
\gamma_i&>&0,\qquad
i=1,\ldots,m-1,\qquad
\mbox{with }\gamma=\sum_{i=1}^{m-1}\gamma_i\geq1
\end{eqnarray*}
and
\[
C=\Biggl\{w_{ij}\geq0\dvt i\wedge j=1, i\vee j\leq m, \sum_{j=1}^mw_{1j}=a,
\sum_{i=1}^mw_{i1}=b\Biggr\},\vadjust{\goodbreak}
\]
then
\begin{eqnarray*}
&&\int_{C}w_{11}^{\alpha-1}\prod_{j=2}^mw_{1j}^{\beta_{j-1}-1}\prod
_{i=2}^mw_{i1}^
{\gamma_{i-1}-1}\,\mathrm{d}w_{11}\prod_{j=2}^{m-1}\mathrm{d}w_{1j}\prod_{i=2}^{m-1}\mathrm{d}
w_{i1}\\
&&\quad\leq
\bigl(B(\alpha,\beta)+
B(\alpha,\gamma)\bigr)\frac{\prod_{j=1}^{m-1}\Gamma(\beta_j)\prod_{
i=1}^{m-1}\Gamma(\gamma_i)}
{\Gamma(\beta)\Gamma(\gamma)}.
\end{eqnarray*}
\end{lemma}

\begin{pf}
Let
\[
K(a,b,\alpha,\beta,\gamma)=\int_0^{a\wedge
b}w^{\alpha-1}(a-w)^{\beta-1}(b-w)^{\gamma-1}\,\mathrm{d}w.
\]
If $a<b$, then
\begin{eqnarray*}
K(a,b,\alpha,\beta,\gamma)&=&\int_0^aw^{\alpha-1}(a-w)^{\beta
-1}(b-w)^{\gamma-1}\,\mathrm{d}w\\
&\leq& b^{\gamma-1}\int_0^aw^{\alpha-1}(a-w)^{\beta-1}\,\mathrm{d}w\\
&=& a^{\alpha+\beta-1}b^{\gamma-1}B(\alpha,\beta)\leq B(\alpha
,\beta).
\end{eqnarray*}
In the same way, if $b<a$, then $K(a,b,\alpha,\beta,\gamma)\leq
B(\alpha,\gamma)$, so that
\begin{equation}\label{Kbound}
K(a,b,\alpha,\beta,\gamma)\leq B(\alpha,\beta)+B(\alpha,\gamma).
\end{equation}
Now, let $W_{11}$ be a random variable on $(0,a\wedge b)$ with density
\[
\frac{1}{K(a,b,\alpha,\beta,\gamma)}w_{11}^{\alpha
-1}(a-w_{11})^{\beta-1}(b-w_
{11})^{\gamma-1},
\]
let $(U_{12},\ldots,U_{1m})$ be a random vector distributed
according to a $\operatorname{Dirichlet}(\beta_1,\ldots,\beta_{m-1})$, let
$(U_{21},\ldots,U_{m1})$ be distributed according to a
$\operatorname{Dirichlet}(\gamma_1,\ldots,\gamma_{m-1})$ and further assume independence
between $W_{11}$, $(U_{12},\ldots,U_{1m})$ and $(U_{21},\ldots,U
_{m1})$. Let $W_{1j}=(a-W_{11})U_{1j},$ $j=2,\ldots,m,$ and
$W_{i1}=(b-W_{11})U_{i1},$ $i=2,\ldots,m.$ From this construction, given
$W_{11}=w_{
11}$, we have that $(W_{12}, \ldots, W_{1m})$ and $(W_{21}, \ldots,
W_{m1})$ are
conditionally independent with conditional densities given,
respectively, by
\[
\frac{1}{(a-w_{11})^{\beta-1}}\frac{\Gamma(\beta)}{\prod_{i=1}^{m-1}
\Gamma(\beta_i)}w_{12}^{\beta_1-1}\cdots w_{1m}^{\beta_{m-1}-1},
\]
with $w_{1j}\geq0$, $j=2,\ldots,m$, $\sum_{2\leq j \leq
m}w_{1j}=a-w_{11}$ and
\[
\frac{1}{(b-w_{11})^{\gamma-1}}\frac{\Gamma(\gamma)}{\prod_{i=1}^{m-1}
\Gamma(\gamma_i)}w_{21}^{\gamma_1-1}\cdots w_{m1}^{\gamma_{m-1}-1},
\]
with $w_{i1}\geq0$, $i=2,\ldots,m$, $\sum_{2\leq i \leq m}w_{i1}=b-w_{11}$.
This construction, together with inequality (\ref{Kbound}), implies
the result,
namely
\begin{eqnarray*}
&&\int_{C}w_{11}^{\alpha-1}\prod_{j=2}^mw_{1j}^{\beta_{j-1}-1}\prod
_{i=2}^mw_{i1}^
{\gamma_{i-1}-1}\,\mathrm{d}w_{11}\prod_{j=2}^{m-1}\mathrm{d}w_{1j}\prod_{i=2}^{m-1}
\mathrm{d}w_{i1}\\
&&\quad \leq
\bigl(B(\alpha,\beta)+B(\alpha,\gamma)\bigr)\frac{\prod_
{j=1}^{m-1}\Gamma(\beta_j)\prod_{i=1}^{m-1}\Gamma(\gamma_i)}{
\Gamma(\beta)\Gamma(\gamma)}.
\end{eqnarray*}
\upqed\end{pf}

\begin{lemma}\label{binomial}
Consider $X$, a $\operatorname{Binomial}(n,p)$ random variable.
We have
\[
\sup_{0\leq p \leq1}
\expec_p[|X-np|]=\cases{
1/B\bigl(1/2,(n+1)/2\bigr), &\quad  if $n$ is odd, \cr
\bigl(1-(n+1)^{-2}\bigr)^{n/2}
\bigl(1+(n+1)^{-2}\bigr)1/B(1/2,n/2), &\quad  if $n$ is even.
}
\]
\end{lemma}

\begin{pf}
Let $\mu_n(p)=\expec_p[|X-np|].$ We have
\[
\mu_n(p)=\frac{2n!}{(\lfloor np \rfloor)!(n-1-\lfloor np \rfloor)!}
p^{\lfloor np \rfloor+1}(1-p)^{n-\lfloor np \rfloor}\qquad
\mbox{for all }n\geq1, p\in[0,1],
\]
where $\lfloor x \rfloor=\max\{n\dvt n\leq x, n \mbox{ is an
integer}\}$ for
all $x$. Therefore,
\begin{eqnarray*}
\sup_{0\leq p \leq1}
\mu_n(p)
& = &
\max_{0\leq k \leq n-1} \sup_{\{p\dvt\lfloor np \rfloor=k\}} \mu
_n(p)\\
& = &
\max_{0\leq k \leq n-1} \mu_n\biggl(\frac{k+1}{n+1}\biggr),
\end{eqnarray*}
and, in particular, $\sup_{0\leq p \leq1}
\mu_1(p)\,{=}\,\mu_1(1/2)\,{=}\,1/2\,{=}\,1/B(1/2,1)$.
Now assume that \mbox{$n\,{>}\,1$}.
Let
\[
\nu_n(k)=
\mu_n\biggl(\frac{k+2}{n+1}\biggr) \Big/ \mu_n\biggl(\frac
{k+1}{n+1}\biggr),
\]
for $k=0,\ldots,n-2$. We have
%
\[
\nu_n(k)=
\frac{(1+(k+1)^{-1})^{k+2}}{
(1+(n-k-1)^{-1})^{n-k}}
\quad\mbox{and}\quad
\nu_n(k)=\frac{1}{\nu_n(n-2-k)}\qquad
\mbox{for }k=0,\ldots,n-2.
\]
However,
\[
\frac{d}{dt}\log\biggl( 1+\frac{1}{t} \biggr)^{t+1}=
\log\biggl( 1+\frac{1}{t}\biggr) - \frac{1}{t} < 0\qquad
\mbox{for all }t> 1.
\]
This implies that
$\nu_n$ decreases on $\{0,\ldots,n-2\}.$
Therefore,
\begin{eqnarray*}
\mu_n\biggl(\frac{1}{n+1}\biggr)&<&\cdots<\mu_n\biggl(\frac{(n+1)/2}{
n+1}\biggr)>\cdots>\mu_n\biggl(\frac{n}{n+1}\biggr)\qquad
\mbox{if }
n\mbox{ is odd,}\\
\mu_n\biggl(\frac{1}{n+1}\biggr)&<&\cdots<\mu_n\biggl(\frac{n/2}{
n+1}\biggr)=\mu_n\biggl(\frac{n/2+1}{n+1}\biggr)>\cdots>\mu_n
\biggl(\frac{n}{n+1}\biggr)\qquad
\mbox{if }
n\mbox{ is even.}
\end{eqnarray*}
The final expression is obtained using the following identity:
\[
n! =2^n\Gamma\biggl(\frac{n}{2}+1\biggr)\Gamma\biggl(\frac{n+1}{
2}\biggr)\Big/\Gamma\biggl(\frac{1}{2}\biggr)\qquad
\mbox{for all }n\geq0.
\]
\upqed\end{pf}
\end{appendix}

\section*{Acknowledgements}

We are particularly grateful to the referee and to the associate editor
for their comments and suggestions that have led to a much improved
version of
this paper. We wish to thank Daniel Stubbs and the staff at R\'{e}seau
Qu\'{e}b\'{e}cois de Calcul Haute Performance (RQCHP) for their
valuable help
with high performance computing. This research was supported by the Natural
Sciences and Engineering Research Council of Canada (NSERC).

%

\printhistory

\end{document}